\documentclass[runningheads]{llncs}
\usepackage[utf8]{inputenc}
\pdfoutput=1
\usepackage{graphicx}
\usepackage{enumitem}
\usepackage{makecell}
\usepackage{multirow}
\usepackage{comment}
\usepackage{amssymb}
\usepackage{multicol}
\usepackage{listings}
\usepackage{url}
\usepackage{pifont}
\newcommand{\cmark}{\ding{51}}
\newcommand{\xmark}{\ding{55}}
\usepackage{hyperref}
\usepackage{todonotes}
\newcommand{\rev}[1]{\textcolor{black}{#1}}

\title{Improving Web API Usage Logging}
\author{Rediana Ko\c ci, Xavier Franch, Petar Jovanovic, Alberto Abell\'o \\Universitat Polit\`ecnica de Catalunya, BarcelonaTech\\{\{koci, franch, petar, aabello\}}@essi.upc.edu}

\date{November 2020}

\begin{document}
\authorrunning\space {R. Ko\c ci et al.}
\maketitle

\begin{abstract}
    A Web API (WAPI) is a type of API whose interaction with its consumers is done through the Internet. While being accessed through the Internet can be challenging, mostly when WAPIs evolve, it gives providers the possibility to monitor their usage, and understand and analyze consumers' behavior. Currently, WAPI usage is mostly logged for traffic monitoring and troubleshooting. Even though they contain invaluable information \rev{regarding consumers' behavior}, they are not sufficiently used by providers. In this paper, \rev{we first consider two phases of the application development lifecycle, and based on them we distinguish two different types of usage logs, namely development logs and production logs. For each of them we show the potential analyses (e.g., WAPI usability evaluation, consumers' needs identification)} that can be performed, as well as the main impediments, that may be caused by the unsuitable log format. We then conduct a case study using logs of the same WAPI from different deployments and different formats, to demonstrate the occurrence of these impediments and at the same time the importance of a proper log format. Next, based on the case study results, we present the main quality issues of WAPI log data and explain their impact on data analyses. For each of them, we give some practical suggestions on how to deal with them, as well as mitigating their root cause.
\end{abstract}
\keywords{Web API, usage logs, log format, pre-processing.}
\section{Introduction}
\sloppy
An increasing number of organizations and institutions are exposing their data and services by means of Application Programming Interfaces (APIs). Different from traditional APIs (i.e., statically linked APIs), which are accessed locally by consumers, web APIs (WAPIs) are exposed, thus accessed, through the network, using standard web protocols \cite{tan2016}. As the interaction between WAPIs and their consumers is done typically through the Internet, both parts end up loosely connected. 

This loosely coupled connection becomes eventually challenging, mostly during WAPI evolution, when as a boomerang effect, consumers end up strongly tight to WAPIs \cite{espinha2015}. If providers release a new version and decide to discontinue the former ones, consumers are obliged to upgrade their applications to the new version and adapt them to the changes. Consequently, WAPIs end up driving the evolution of their consumers' application \cite{espinha2014}, \cite{eilertsen2018}. Knowing the considerable impact WAPIs have on their consumers, providers would benefit from consumers' feedback to understand their needs and problems when using the WAPI~\cite{murphy2018}. 

Currently, API providers face a lot of difficulties in collecting and analyzing consumers' feedback from several sources (often informal ones), e.g., bug reports, issue tracking systems, online forums and discussions. \cite{zhang2020}. Furthermore, feedback collection and analysis turns out to be expensive in terms of time, thus difficult to scale. Actually, in the WAPI case, this feedback can be gathered in a more centralized way.

\begin{figure}
    \centering
    \vspace{-20pt}
    \includegraphics[width=\textwidth]{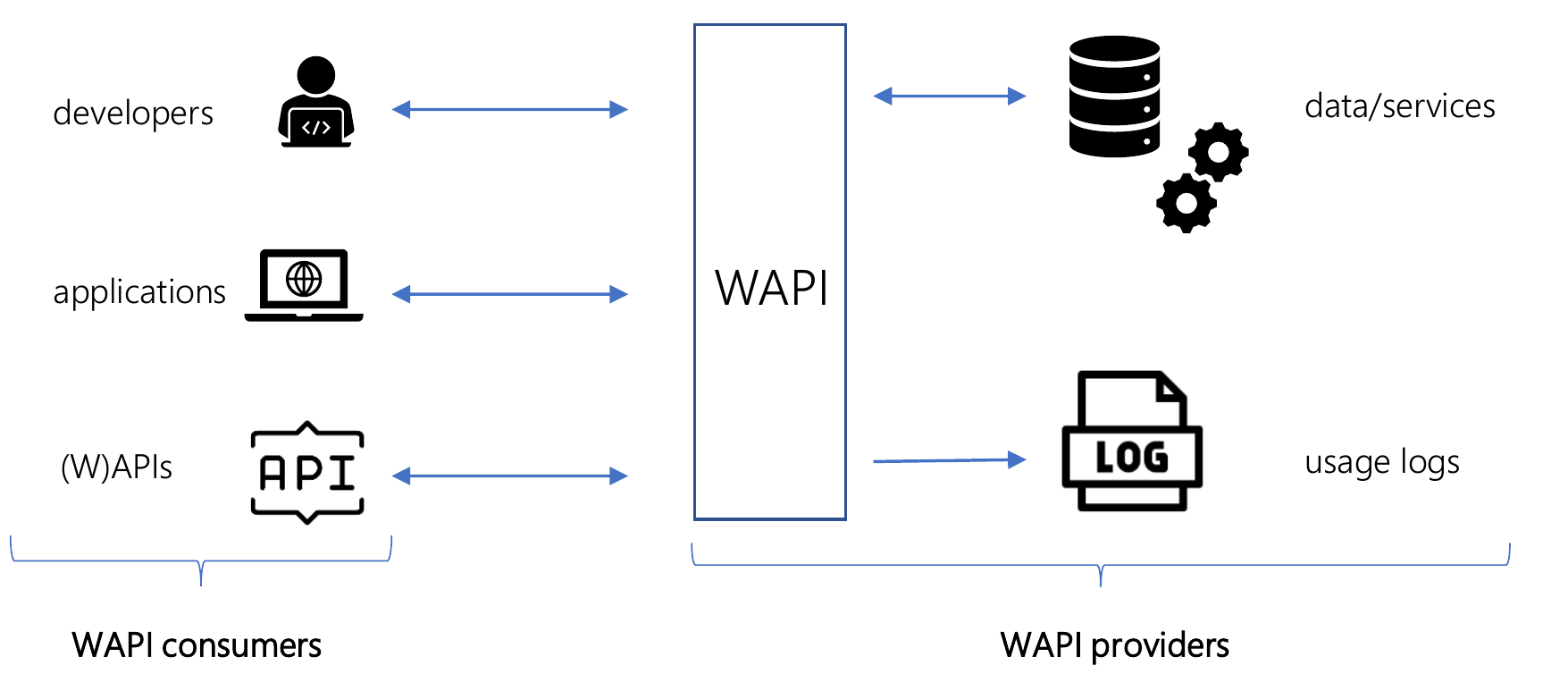}
    \vspace{-20pt}
    \caption{\rev{The interaction between consumers, providers and WAPIs}}
    \vspace{-20pt}
    \label{traffic}
\end{figure}

While being accessed through the network poses some challenges for consumers, it enables providers to monitor the usage of their WAPIs, by logging every request that consumers make to them (see Fig. \ref{traffic}). WAPI usage logs, besides coming from a trustworthy source of information, can be gathered in a straightforward, inexpensive way, \rev{and completely transparent to the WAPI consumers.}

Currently, WAPI usage logs are mostly being used to feed monitoring tools. These tools typically provide automatic alerts when WAPI endpoints fail, and reporting dashboards that visualize several performance metrics \cite{doerrfeld}. Since these logs are not designed to be further analyzed with regards to consumers' behavior, they may lack some critical information, like identifiers for consumers' applications, etc. Moreover, they are complex (e.g., unstructured, high volume data) and somewhat noisy. Applications' design and the way their end users use them can veil interesting and important WAPI usage patterns. Therefore, understanding consumers' behavior and inferring their needs from WAPI usage logs becomes \rev{essential for providing high quality WAPIs, tailored to the real needs of their consumers.}

This paper is building upon our previous work \cite{koci2020}, where we measured the usability of WAPIs by analyzing their usage logs generated during the development phase of consumers' applications. Based on the challenges faced while working with WAPI usage logs, and the surprising lack of attention this topic (i.e., WAPI usage logs analysis) had gained, we saw convenient to summarize our experience and research in the field, into a set of practical suggestions to enhance the logging of WAPI usage for more specialised analyses. 

To this end, the contribution of this work is fourfold. We focus first (i) on showing the potential and then (ii) the main impediments of proactively using WAPI usage logs regarding consumers' behavior. We then conduct a case study using logs of the same WAPI from different deployments, using different log formats, (iii) to show the importance of logs' structure and content in preparing the data for further analyses. Next, based on the case study findings, the analysis requirements, WAPI logs structure, and also after reviewing the relevant literature, (iv) we classify and define the main issues and obstacles that hinder the application of various analyses on the logs. For each of the issues, we describe the impact their occurrence may have on the analyses result, and propose mitigation actions.

The remainder of this paper is organized as follows. In Section \ref{rwork}, we report on the current use of WAPI usage logs and practices to pre-process and deal with their quality issues. In Section \ref{motivation}, we give the main motivation behind our work. We introduce the WAPI logs structure and content, and based on them we propose some purpose-specific analyses that can be applied on these logs. In Section \ref{format}, we introduce two pre-processing challenges, whose accuracy may be affected by the format of the logs. In Section \ref{casestudy}, we introduce our case study and the experiments we performed. In Section \ref{issues}, we provide the set of WAPI quality issues inferred from the case study, and discuss their impact and mitigation. In Section \ref{conclusion}, we conclude the paper and present some ideas for future work. 


\section {Related Work} \label{rwork}

Few works build their analysis on WAPI usage logs comprised by the URL requests made to the WAPI \cite{suter2015}, \cite{koci2020}, \cite{macvean2016}. Thus, not only these logs' potential is still unrevealed, but even quality issues related to them or their pre-processing and preparation have not gained the deserved attention. 

For instance, Suter and Wittern \cite{suter2015} used the usage logs to infer WAPI description (the endpoint structure and
parameters) from them. They reported that the results of their methods were impeded from the incomplete and noisy nature of these log data. In our previous paper \cite{koci2020}, we proposed an approach to measure the usability of WAPIs by analyzing the usage logs generated during the development phase of consumers' application. We described the pre-processing steps of WAPI usage logs in general, and then demonstrated how we dealt with specific obstacles in the log data from the case study (e.g., data structuring, generalization). Macvean et al. \cite{macvean2016} analyzed WAPI usage logs from Google API Explorer, and generated from them several structural factors (e.g., number of parameters, number of methods) to study their usability. Even though they showed the potential of analyzing these logs, they did not tackle the quality issues related to them or challenges during their preparation.

On the other hand, web log preparation and pre-processing are widely studied, as part of web log analysis, extensively applied regarding usage and usability of software and web applications \cite{goel2013}. We mention below some of these works, as some WAPI logs quality issues are similar to the web logs ones. 

One of the most discussed issues of log pre-processing is session identification. A session represents the interaction of a user with a website within a time frame (usually expires after a certain amount of time of inactivity). There exist several heuristics for reconstructing sessions, mostly coming from web mining applications \cite{srivastava2019}, \cite{srivastava2000}, \cite{tanasa2004}, \cite{kapusta2014}, \cite{berendt2001}, \cite{spiliopoulou2003}. Spiliopoulou et al. \cite{spiliopoulou2003} applied different heuristics (total session duration, page-stay duration, etc.) to reconstruct the sessions from the server log data. They evaluated the performance of these heuristics to the server log of a university site by comparing the reconstructed sessions with the real ones. Their experiments showed that there is no one best heuristic for all cases, and it depends on site's structure and traffic. Kapusta et al. \cite{kapusta2014} analyzed the logs from a commercial bank portal to identify the users' sessions. They applied different time window thresholds and heuristics, and based on the usefulness of the rules extracted in each case, they evaluated the best threshold. Tanasa and Trousse \cite{tanasa2004} described in detail all the steps of log pre-processing (i.e. data fusion, data cleaning, data structuring and data generalization), indicating the most challenging issues and how to overcome them. They pointed out the importance of pre-processing in data analysis effectiveness, and among others, agreed on the need for a better log systems. 

The above mentioned works focused on general web usage logs, thus the concerns raised were related to the analyses applied on them (e.g., to identify users navigation behavior in order to predict their next actions, to evaluate software design usability, to monitor the traffic for performance reporting). While some of the issues of working with general web logs are similar to WAPI usage logs, the latter pose some added challenges, related to the requirements of the specific analyses that can be applied on them, as well as the WAPI design. 

Bose et al. \cite{bose2013} focused their work on the requirements of process mining (an analyzing technique that can be applied on log data) and log data quality issues that may affect its results. They presented 4 process characteristic issues and 27 event log quality issues that hinder the applicability of several process mining techniques and affect the results' quality, but did not provide solutions on how to address those issues. Along similar lines, Suriadi et al. \cite{suriadi2017} described a set of data quality issues, frequently found in process mining event logs. Based on their experience in performing process mining analyses, they introduced 11 event log imperfection patterns, which can be used in several domains. While both of the works are too specific for process mining requirements, they refer to event logs in general in terms of the domain. Thus, some of the issues introduced cannot be applied in the WAPI domain (e.g., issues related to the manual data entry). 

In this paper, driven by the lack of attention WAPI usage logs analysis has gained, we show several potential analyses that can be applied on them, mostly based on their nature and content. We discuss the posed challenges in analyzing these logs and eliciting the needed information, followed by recommendations on how to better log the WAPI usage, which until now, remain quite unexplored.
\section{The potential of WAPI usage logs} \label{motivation}

Providers typically log their WAPIs' usage by recording all requests done against the WAPIs. Every time a consumers' application issues a request to a WAPI, a log entry is generated and stored in the usage log file. The information that is logged for each request and its format may vary due to different logging system setting parameters and also providers' decision about logs design. For example, using the Apache Custom Log Format\footnote{\url{http://httpd.apache.org/docs/current/mod/mod_log_config.html}} (a flexible and customizable format), when an application makes a request like \url{https://maps.googleapis.com/maps/api/distancematrix/json?origins=MNAC&destinations=MACBA&mode=driving&key=API_IDENTIFIER} (in the form of a URL) to a WAPI endpoint of the Google Maps Platform, the following information could be logged by providers (based on the logging system configuration): the IP address of the application's user, the time when the consumer made the request, the request body that contains the request method (GET), path (\url {maps.googleapis.com/maps/api/distancematrix/json}) and query (\url{origins=MNAC&destinations=MACBA&mode=driving&key=API\_IDENTIFIER}), the protocol (HTTP/1.1), the time needed to respond to the request, the status code (e.g., 200 if the request was successful), the size of the object returned, the address of the page that initiated the request, information about the operating system or browser used, and other fields comprising different aspects of consumers' applications and their users. 

\begin{figure}
    \centering
   \vspace{-10pt}
    \includegraphics[width=\textwidth]{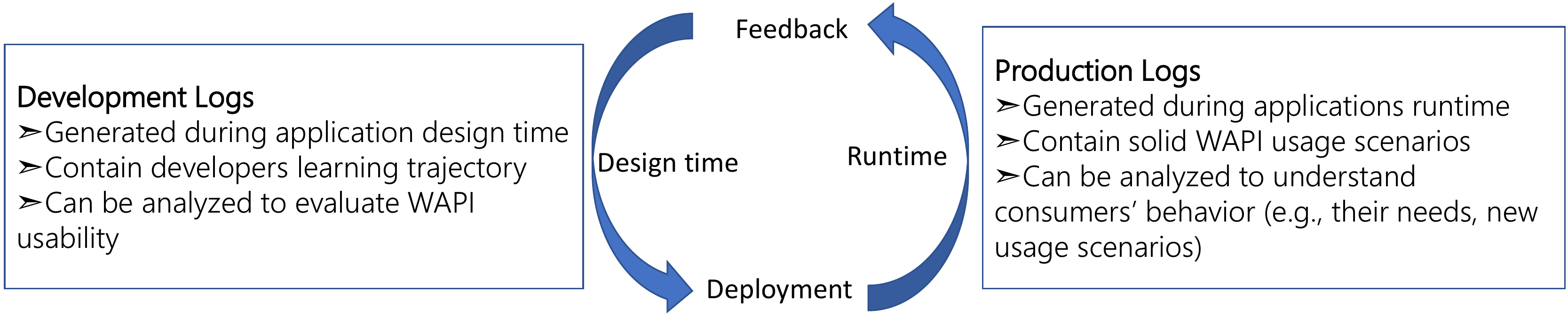}
   \vspace{-20pt}
    \caption{The development lifecycle of consumers' application}
     \vspace{-20pt}
    \label{logs}
\end{figure}

We can think of usage logs as traces that consumers leave when using the WAPI. If the information in these traces is analyzed in the proper way, it can reveal useful knowledge. They show which endpoints the consumers have accessed, in which order, with which frequency, and with which parameters. As applications are the actual WAPI consumers, we should consider the different ways they consume WAPIs over their own lifecycle. Basically, applications interact with the WAPIs during design time and runtime, over both of which they manifest different aspects of their behavior. Following on from this, we distinguish two types of logs: (i) development logs, and (ii) production logs (Figure~\ref{logs}).

Development logs are generated at design time, while developers build and test their applications. During this phase, they make the first integration of the WAPI into their applications, or implement new developed features. Thus, these logs show their attempts in using the WAPI, the endpoints they struggle more with, specific mistakes they do while using and learning the WAPI, etc. \cite{koci2020}, \cite{macvean2016}. By analyzing these logs, providers may evaluate the usability of their WAPI from the consumers' perspective. For instance, they may decide to change the name of elements (endpoints, parameters) consumers have difficulty in learning or memorizing, improve the documentation for endpoints that seem not clear to consumers, detail the error messages when they detect that consumers are repeating continuously the same errors without understanding how to fix it, etc. 

On the other hand, production logs are generated during applications runtime, while application are being used by end users. Since the applications are released for public use, it is assumed that they are quite steady, without erroneous WAPIs requests. WAPI requests are predetermined by the implemented functionalities of the applications, different from the development phase, where developers may freely try different requests, several times, and pose any query. Indeed, production logs contain real and solid WAPI usage scenarios, the right order in which developers make the requests to WAPI to achieve specific goals, different workarounds created to accomplish tasks for which there are no WAPI endpoints developed, or the actual frequency of certain requests or sequences of requests, that show the real consumption of WAPIs. By analyzing these logs, providers may identify consumers' needs for new features, and implement the corresponding endpoints. These logs may reveal new usage scenarios providers may have not thought about before, instructing them in including these scenarios in the documentation. Besides these, providers may identify ways of improving the WAPI based on how consumers use it, merging endpoints that are always called together for a specific purpose, or creating new endpoints, derivative from the ones that are always called with specific values for some parameters. 

\rev{Even though both types of logs provide useful information about WAPI consumption and perception from consumers, preparing and analyzing them is arduous. First, it is not always trivial distinguishing these logs from each other, as they often are stored together in the same files. Secondly, consumers' applications design and the way users interact with them will be manifested in the production logs, obfuscating the inference of the real WAPI usage patterns. Providers should identify the patterns that represent real usage scenarios, from the ones deriving from applications design and users flow. Finally, as providers store these logs typically for traffic monitoring, they do not consider the requirements that specific analyses may have. Thus, unawarely, they may neglect the importance of the log format, and even leave out crucial information for consumers' identification, adversely affecting not only the analysis results, but also the logs pre-processing.} 
\section{How does the logs format affect the pre-processing?} \label{format}
\begin{figure}
    \centering
    \vspace{-20pt}
    \includegraphics[width=\textwidth]{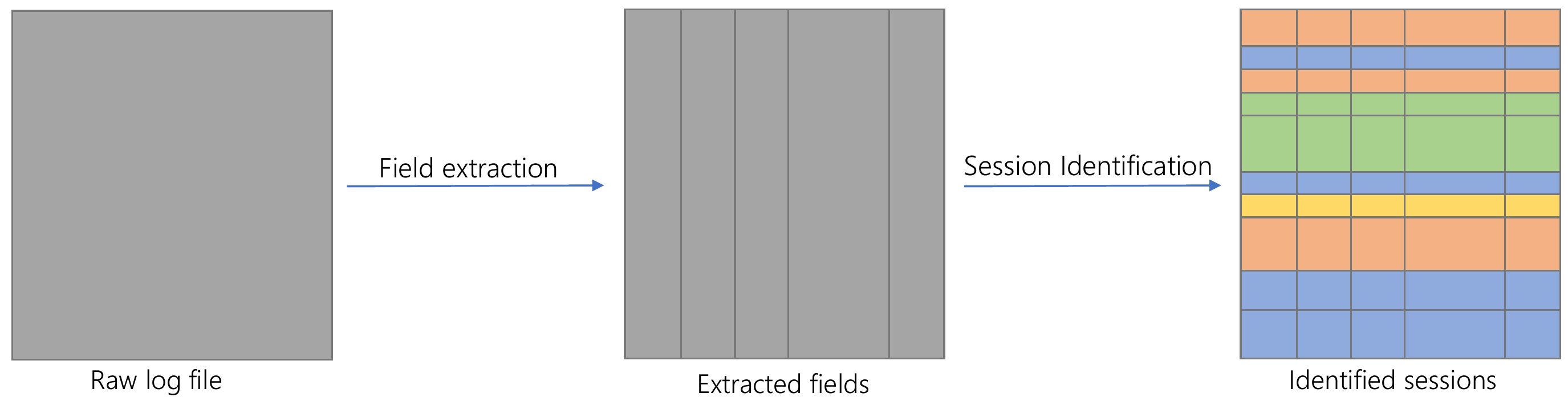}
    \caption{Field extraction and session identification}
    \vspace{-20pt}
    \label{slide}
\end{figure}

\rev{The pre-processing phase is typically counted as the most difficult and time-consuming part of log analysis \cite{srivastava2019}. It basically consists of four main steps: (i) data fusion, consisting in gathering and merging log files from different sources, (ii) data cleaning, consisting in removing irrelevant data and completing missing values, (iii) data structuring, consisting in segmenting the log file in users' sessions, and (iv) data generalization, consisting in generalizing the dynamic part (i.e., parameter values) of requests \cite{tanasa2004}, \cite{koci2020}. In this section we will cover two challenges from WAPI usage logs pre-processing, namely field extraction from data cleaning phase, and session identification from data structuring, as the two challenges of pre-processing that are directly affected by the log format and the way the usage is being logged (Figure \ref{slide}).} 

\rev{\textbf{Field extraction.} Usage logs are stored in text files. Each log entry contains several fields, each containing specific information. Field extraction consists exactly in the separation of the log entry in several fields. It is typically performed right before data cleaning, so that log entries can be filtered based on the value of their specific fields (e.g., request method, request body).}
For example, the following log entry should be transformed from a single string into the set of fields it contains: \texttt{127.0.0.1 - - [24/Jun/2019:20:22:26 +0000] GET /api/29/system/info HTTP/1.0 200 891 https://.../dhis-web-dashboard/index.html Mozilla/5.0 (Windows NT 6.1; Win64; x64) AppleWebKit/537.36 (KHTML, like Gecko) Chrome/75.0.3770.100 Safari/537.36}, extracted fields:
\vspace{-5pt}
\begin{lstlisting}[label={fields}, basicstyle=\ttfamily]
Client IP  : 127.0.0.1  
Timestamp  : 24/Jun/2019:20:22:26 +0000
Request    : GET /api/29/system/info HTTP/1.0 
Status code: 200        
Object Size: 891 
Referer    : https://.../dhis-web-dashboard/index.html 
User-Agent : Mozilla/5.0 ... Safari/537.36
\end{lstlisting} 
\vspace{-5pt}
\rev{There are several ways to parse log files information, including regular expressions, predefined parsers, custom grok parsers (pattern matching syntax used by ElasticSearch). Providers should decide on a log format that can be easily parsed, in order to enable a simple querying of fields value.}   

\rev{\textbf{Session identification.} This challenge (often called \emph{sessioning}), refers to grouping together into the same session, all the log entries (i.e., requests) coming from each user during the time frame of a visit, trying not to leave out any log entry, as well as not assigning wrong ones. It is one of the main issues with WAPI usage logs. Most of the WAPIs are stateless, meaning that the server does not store the state, thus no sessions are generated. The lack of sessions' identifiers may seriously impede the applicability of several analyses. For instance, one of process mining requirements is for event logs to have case identifiers, which assign each log entry to a specific case \cite{aalst2016}. Session identifiers must be inferred combining other available information in the logs. Sessioning heuristic is a method for constructing sessions based on assumptions about users’ behavior or the site/application characteristics. Two of the most applied methods are time-based heuristics and navigation-based heuristics \cite{berendt2001}. As both of them are built under the hypothesis of an already launched and ready to be used application, these heuristics apply only to the production logs.}

\begin{itemize}[leftmargin=*]
    \item \textit {Time-based heuristics} construct the sessions based on either: (i) the duration of a user's entire visit to the application, which should not surpass a maximum threshold $\delta$, typically taken 30 minutes \cite{berendt2001}, or (ii) the time a user spend on one page of the application (i.e., page-stay heuristic), which should not surpass a maximum threshold $\theta$, defined based on pages average contents and application nature.
   
    \item \textit{Navigation-based heuristics} construct the sessions based on the assumption on how the applications' pages are related. The rationale behind is that users' navigational flow in the application is predetermined since its implementation. For native (or desktop) applications this flow is fixed. For web ones, which are accessed through a browser, users rarely type themselves the URL of a page, but rather follow the hyperlinks and the navigation bar. In the usage logs, the information about the page initiating the actual request is contained under the referer field. This field can have a null value (``-") when the users type the request directly in the browser, or when an application is first opened. 
\end{itemize}
\section{Case study} \label{casestudy}

We perform field extraction and session identification in order to demonstrate the importance of specific fields of the log format, and the impact their lack may cause to both of these challenges. We conduct a case study using logs of the District Health Information Software 2 (DHIS2) WAPI. DHIS2 is an open source, web-based health management information system platform used worldwide from various institutions and NGOs for data entry, data quality checks and reporting. It has an open REST WAPI, used by more than 60 native applications. External software can make use of the open API, by connecting directly to it or through an interoperability layer.

DHIS2 is instantiated as World Health Organization (WHO) Integrated Data Platform\footnote{\url{http://mss4ntd.essi.upc.edu/wiki/index.php?title=WHO\_Integrated\_Data\_Platform\_(WIDP)}} (WIDP), and is used by several WHO departments for routine disease surveillance and country reporting. For the analysis, we use the production logs from WIDP, and from Médecins Sans Fontières (MSF), another DHIS2 instance used for field data collection and as a central repository for medical data.

\rev{Both of these instances use the same DHIS2 WAPI and the same set of applications accessing it.} But, being deployed and used independently, the logs coming from them have different formats, providing us with different information that we can use to structure and prepare the logs for further analyses (Table \ref{formats}). 

    \begin{table*}[htbp]
    \vspace{-10pt}
     \caption{Log formats of the two DHIS2's deployments under study}
        \begin{center}
      \begin{tabular}{ |c|c|c|c|c|c|c|c|c| } 
        \hline
         \makecell{Deployment} & \makecell{Client IP \\ address} & \makecell{Timestamp \\ granularity} & \makecell{Duration} & \makecell{Request} & \makecell{Status \\ Code} & \makecell{Object \\ Size} & \makecell{Referer} & \makecell{User \\ Agent}\\ 
         \hline
         MSF & \xmark & Second & \cmark & \cmark  & \cmark  & \cmark  & \cmark  & \cmark  \\ 
         \hline
        WIDP  & \cmark  & Millisecond  & \cmark  & \cmark  & \cmark  & \cmark  & \xmark  & \cmark  \\  
         \hline
        \end{tabular}
        \end{center}
        \vspace{-20pt}
        \label{formats}
        \end{table*}
\begin{enumerate}[leftmargin=0pt]

\item \textbf{Field extraction.} We perform field extraction by using regular expressions in JAVA. Request body, referer and user-agent, are the parts that generate more errors while parsing, as they may include spaces and special characters, sometimes used to separate the fields. We show the example of the user-agents values in WIDP log data, which typically have in their body comma, semicolon, and spaces: Mozilla/5.0 (Windows NT 10.0; Win64; x64) AppleWebKit/537.36 (KHTML, like Gecko) Chrome/80.0.3987.122 Safari/537.36. We have to perform some extra manual work to handle the errors, like splitting these fields into several parts, and then joining them, without cluttering parts of different fields.
\item \textbf{Session identification.}
We apply the page-stay heuristic and the navigation one combined with time constraint. We perform the experiments on log data from MSF, since the log format contains information about the referer (Table~\ref{formats}), required in navigation-based method. Since we do not (and cannot) have the data with the real and correct session identifiers to evaluate the performance of the resulting techniques, we assess the correctness of the constructed sessions based on different statistics. The results of the analysis are shown in Table \ref{heuristics}.

The WAPI log file from MSF has requests from different applications installed in the platform, used by different users. Our first concern consists in eliciting the user and the application that submitted the request. MSF uses a proxy server, thus the information under the client IP can not be used in user identification. Also, the log entries do not have information about which application submitted each request. The only information available are the log entries specifying the opening of an application: ``{\fontfamily{qcr}\selectfont GET /\{nameOfTheApplication\}/index.action}''. 
\begin{itemize}[leftmargin=0pt]
\item \textit{Time-based heuristic.} We start with the time based heuristic. Even though the aim of the experiments is not to find the best threshold value, we perform the experiments for two different thresholds (5 and 15 minutes), to see if their values make any significant changes in the sessioning of the log entries. If the timestamps of the requests after an application opening have a difference of less than the defined threshold, then they are considered part of the same session. Otherwise, they are discarded, as we cannot know the application making the requests. Using the example in Listing \ref{timeHeuristicMSF}, the requests after the opening of App2 are considered part of the same session if $t_4-t_3 \leqslant 5 min (15min)$ and $t_5-t_4 \leqslant 5 min (15 min)$, or discarded otherwise. On the other hand, these requests may comply with the threshold for App1 as well ($t_4-t_2 \leqslant 5 min (15min)$, $t_5-t_2 \leqslant 5 min (15 min)$). As a result, requests that may come from App1 session, may be assigned to the next session, possibly increasing this way the error rate in two directions: not including the right log entries in Session1, and including wrong ones in Session2. 
 
\begin{lstlisting}[caption={Time based heuristic on MSF logs},label={timeHeuristicMSF}, basicstyle=\ttfamily]
1. GET App1/index.action        t$_1$  Session1
2. request from App1            t$_2$  Session1
3. GET App2/index.action        t$_3$  Session2 
4. request from App1 or App2    t$_4$  Session1 or Session2 
5. request from App1 or App2    t$_5$  Session1 or Session2 
   \end{lstlisting} 
   
As seen from the situation in the log snippet in Listing \ref{timeHeuristicMSF}, regardless of the timestamps, we cannot be sure about the session of the requests after two applications open simultaneously. It is worth pointing out that the uncertainty would still persist, even if the information about client IP address was in the log file, as the same user could open several applications at the same time. 
 
We can see in Table \ref{heuristics} (first two rows), the overall number of sessions (for sessions with more than 3 requests) in the data, the average sessions' duration, and the average sessions' size (number of requests in a session). We are not further analyzing these metrics, to see which of the thresholds is more performative, as from the interpretation, the time threshold, used in isolation from other fields, is not enough for identifying sessions. \rev{But we will compare them with the metrics derived from the results of the navigational-based heuristic.} 
    
\begin{table*}[htbp]
  \vspace{-10pt}
  \caption{Statistics for the defined sessions}
  \begin{center}
    \begin{tabular}{|l|r|r|r|}
      \hline
      \textbf{{Heuristic}}& \textbf{{No. of sessions}}& \textbf{{Avg. duration}}& \textbf{{Avg. size}}\\
     \hline
    time { }5 min& 15,804 & 110 sec & 63 \\
     \hline
    time 15 min& 15,937 & 127 sec & 63\\
     \hline
    time { }5 min, navigation& 8,233& 266 sec & 122\\
     \hline
    time 15 min, navigation& 6,586& 413 sec& 152\\
      \hline
    \end{tabular}
  \end{center}
    \vspace{-20pt}
  \label{heuristics}
\end{table*}

\item \textit{Navigational-based heuristic, combined with time constraint.} Next, we reconstruct the sessions using not only the time difference between the requests, but also the referer. We use this heuristic for two different timeout thresholds. Extending the same example with referer information (see Listing \ref{navigation1MSF}), we can see that the session identification accuracy is straightforward when different applications are being consumed (from the same user or different ones). Log entries 4 and 5 are assigned to the right session, due to the information provided by referer. (The value of refA is ignored, as the request body itself gives the application in use.)
  
 \begin{lstlisting}[caption={Navigation based heuristic on MSF logs, different application},label={navigation1MSF}, basicstyle=\ttfamily]
1. GET App1/index.action   t$_1$  refA  Session1 
2. request from App1       t$_2$  App1  Session1
3. GET App2/index.action   t$_3$  App1  Session2 
4. request from App2       t$_4$  App2  Session2 
5. request from App1       t$_5$  App1  Session1 
   \end{lstlisting} 

We cannot say the same when the same application is being used by different users (see log snippet in Listing \ref{navigation2MSF}). Log entries 4 and 5 may belong to Session1 or Session2, but they may end up in the wrong session, due to the lack of client IP address information.

 \begin{lstlisting}[caption={Navigation based heuristic on MSF logs, same application},label={navigation2MSF}, basicstyle=\ttfamily]
1. GET App1/index.action   t$_1$  refA Session1 
2. request from App1       t$_2$  App1 Session1
3. GET App1/index.action   t$_3$  App1 Session2 
4. request from App1       t$_4$  App1 Session1 or Session2 
5. request from App1       t$_5$  App1 Session1 or Session2 
   \end{lstlisting}  
     \vspace{-10pt}
   \end{itemize}
\end{enumerate}

As seen from Table \ref{heuristics} (two last rows), the metrics from the navigation method are significantly different from the ones when only the time heuristic was used (two first rows). We can see that after using the referer information, for the same logs we have less number of sessions, but larger ones in terms of number of requests and session total duration. This means that, when using only the time heuristic, we are over-splitting the sessions, thus potentially loosing sequences of requests. Besides this, the new sessions created, likely contain mixed requests from different users and different applications, thus possibly creating fake sequences of requests. 

\rev{Even though we applied grounded methods in pre-processing the logs, we admit that challenges like session identification and field extraction may still remain due to the lack of important information in the logs or the format being used. These problems are hard to deal with, and the best way to address them is mitigating the root cause.}

\textit{Assessment.} The performance of the heuristics could be evaluated by comparing the constructed sessions with the real ones. In WAPI usage logs we cannot have the real sessions. Thus, in order to evaluate the accuracy of the reconstructed sessions of logs from MSF, we compare them and the reconstructed sessions of logs from WIDP, in the context of four specific applications. Different from MSF, the logs from the WIDP have information about the client IP addresses, but not the referer (Table \ref{formats}). Consequently, we reconstruct the sessions using client IP and timeout (15 minutes). Then for each application, we extract the distinct requests assigned on both instances (Table~\ref{sessions}). We saw that when using only the time heuristic in MSF, even though the sessions are in average shorter in terms of number of requests (Table \ref{heuristics}), too many distinct and different requests are assigned to each application. The same happen with WIDP, whose logs do not have information about the referer. For each application we explore in detail the distinct requests assigned to them, for all the sessions. We saw that there were WAPI requests, that even though not related to the applications, were assigned to them because of the missing information in the logs. 

\begin{table*}[htbp]
 \vspace{-10pt}  
  \caption{WAPI requests assigned to four applications installed in MSF and WIDP}
  \begin{center}
    \begin{tabular}{|l|r|r|r|}
      \hline
      \makecell{\textbf{Application}}& \makecell{\textbf{MSF (time) }}& \makecell{\textbf{MSF (time, navigation) }}& \makecell{\textbf{WIPD (time)}}\\
     \hline
    dhis-web-event-capture & 106 & 28	& 177 \\
     \hline
    dhis-web-event-reports &	139 &	25	& 118 \\
     \hline
    dhis-web-tracker-capture &	202 &	48 & 124 \\
     \hline
    HMIS-Dictionary	& 136 &	31 & 57 \\
      \hline
    \end{tabular}
  \end{center}
    \vspace{-30pt}
  \label{sessions}
\end{table*}

\rev{As the examples show and the experiments' results support, not being able to identify the users and the applications submitting the WAPI requests, greatly affects the correctness of sessions' identification. We can see an improvement when the referer information is available in the logs, but its exploitation comes with extra pre-processing efforts. } 

\section{Common WAPI logs issues} \label{issues}

In this section, deriving from the case study, we introduce the main WAPI usage logs quality issues, that are responsible for the problems surged during field extraction and session identification. Some of them are related to the nature of WAPI usage logs, hence providers should be aware and consider them before analyzing the logs. Others originate from the way WAPI usage is logged, thus can be eliminated or ameliorated.
\begin{enumerate}[leftmargin=0pt]
\item \textbf{Field extraction}
\begin{itemize}[leftmargin=0pt]
\item Fields' separators part of the fields' body. 

 \textit{Description}: Even though it is recommended to use more human-readable formats for logs, keeping them machine processable is also important. Each log entry consists of several fields, typically separated by specific characters, e.g., comma, semicolon, space. With the help of these separators, providers can perform the fields' extraction. Problems may arise when the fields themselves contain in their body the characters used as separators. The most heterogeneous fields are the request, the referer, and the user agent. 
    
\textit{Impact}: Not addressing this issue may result in extra added time and effort in log data pre-processing. If the special separators are part of the fields body, the automation of field extraction will generate errors, and providers should perform manual work to fix the issue.   
    
\textit{Mitigation}: To facilitate the field extraction, it is recommended to double-quote the fields that might have special characters like request, referrer, user agent, etc. The use of machine parseable formats will increase the automation of pre-processing, and therefore its correctness.
\end{itemize}
\item \textbf{Session identification}
\begin{itemize}[leftmargin=0pt]
    \item Insufficient fields. 
     
     \textit{Description}: WAPI usage logs are usually not logged for the purpose of analyzing consumers' behavior and getting indirectly their feedback. Thus, they often suffer from missing crucial fields for the application of several analysis, or other fields whose presence may enrich the analyses with new insights. We encountered this issue with the log data from WIDP, whose format did not include the referer.
     
     \textit{Impact}: The lack of specific fields may become an impediment for applying several analyses or may affect the accuracy of the analyses' results. The \textit{referer} header is a field that contains the address of the page that made the request. Even though this is an optional field, it contains an important and helpful information to reconstruct the sessions. As demonstrated in the introduced case study, in cases where: (i) the session identifiers are not present, (ii) the client IP address is actually a proxy address, and (iii) session timeout differs between several consumers' applications, the information under referer will help the analysts to better identify the session of a log entry and reconstruct the requests' sequence. Even though providers cannot fully rely on the referer information (it is not in the log when consumers type themselves the request in the browser), it increases the correctness of assigning every log entry to its own session. The \textit{user agent} field, which contains information about the browser and the operating system used when making the request, may as well play an important role in distinguishing the requests coming from different users, thus it should be included in the log format.
     
     \textit{Mitigation}: In order to strike a balance between not leaving out important fields, and at the same time not to log too many fields, providers should decide beforehand on the analyses they will perform on the usage data, and the question they seek to answer. The specific requirements of the analyses should help them in making the right decision. Nevertheless, regardless of the type of analyses, providers should be able to identify different users and applications, and log the needed fields accordingly. 

    \item Missing applications' identifiers. 
    
     \textit{Description}: Applications' identifiers are unique identifiers that providers generate for their consumers, usually to monitor their usage for billing purposes. Consumers must include their applications' identifiers in each WAPI request, so that providers can track their WAPI usage. Currently this practice (of providing application identifiers) is typically followed by providers that have monetized their WAPIs. However it can be used for more than just correctly charging consumers. \rev{We are not covering this in the `Insufficient fields' issue, as more than a field to be included in the log format, it is related to providers decision to generate this kind of identifiers. Consumers may then submit the identifiers as HTTP header, as a query parameter, or as a request body field. }
    
    \textit{Impact}: The lack of applications' identifiers is not likely to impact the accuracy and correctness of analysis results, but it affects the evaluation and the prioritization of the found usage patterns. Suppose that providers will find in the logs specific usage patterns that may indicate the need for some changes in the WAPI. Not knowing which applications are making the requests, providers cannot be sure whether the patterns found are coming from several applications, or from few applications with a lot of users. In this situation, if they decide to perform the indicated changes, they will not know how many applications these changes will affect. To make informed decisions about the implementation of the prescribed changes, they should have the information about these identifiers. Furthermore, under the conditions where sessions' identifiers are missing, application identifiers will help providers in improving the sessioning of the usage logs. We faced this issue with logs from both MSF and WIDP, as DHIS2 provider was not generating these identifiers.  
    
    \textit{Mitigation}: Providers can address this issue by generating unique identifiers for each consumers' application, so that consumers include them in all the requests made to the WAPI. Additionally, in order to differentiate between usage logs created during development/testing phase and production phase, providers should generate different identifiers for each of the phases. As already explained, these usage logs manifest different aspect of consumers' behavior. Thus, providers should be able to separate them in order to accurately apply purpose-specific analysis.    
        
\item Hidden client IP address. 

 \textit{Description}: The client IP address gives the IP addresses of the applications' users. Combined with other information (session timeout, referer, user agent, etc.) this information can be used in users identifications, as well as sessioning. However, if the consumers are using proxy servers, as in the case of MSF, the IP address that appears in the usage logs will not be of the original user doing the request, but that of the proxy server address. As a result, different users may appear under the same client (proxy) IP address in the logs, misleading the user identification process.  We want to note that, as we are analyzing the way applications are consuming the WAPIs, our interest on users' identifications is limited to their help in sessions' identifications. By this means, if the same users appear with different IP addresses each time they use an application (dynamic IP addresses), this will not affect the analysis results.
    
\textit{Impact}: Not being able to distinguish the requests from different users may produce mixed up sequences of requests. The impact can be even more severe if other identifiers (e.g., application identifiers) are also missing in the logs, as in the example in Table \ref{ipaddresses}. 
     \begin{table*}[htbp]
     \vspace{-10pt}
      \caption{Log entries from different users with the same IP address (proxy address).}
     \begin{center}
      \begin{tabular}{ |c|c|c|c| } 
      \hline
         \makecell{{Client IP}} & \makecell{{Request}} & \makecell{{Timestamp}} & \makecell{{Referer}} \\ 
         \hline
         IP1 & request1 & 18/Dec/2020 09:35:27.723 & app1 domain\\ 
         \hline
        IP1 & request2 & 18/Dec/2020 09:35:28.112 & app1 domain\\  
         \hline
         IP1 & request3 & 18/Dec/2020 09:35:33.009 & app2 domain\\
         \hline
          IP1 & request4 & 18/Dec/2020 09:36:07.545 & app3 domain\\
         \hline
          IP1 & request5 & 18/Dec/2020 09:36:36.225 & app1 domain\\
         \hline
        \end{tabular}
        \end{center}
          \vspace{-20pt}
        \label{ipaddresses}
        \end{table*}
        
\textit{Mitigation}: WAPI providers cannot control or fix this issue. Thus, it is important for them to be aware of this problem and to not fully rely on this field for users and sessions' identification. Instead, they should make sure to include other fields in the logs (e.g., referer, user agent), that will help them in better structuring the logs. This was the case of usage logs from MSF. While the client IP addresses were not usable, the referer information helped in logs sessioning.

\item Timestamp coarse granularity. 

 \textit{Description}: The timestamp field shows the exact time the request was made to the WAPI. Even though the logging system stores the timestamp when the request was made, the log entry that represents that request is printed in the log file after the WAPI sends the response to the consumers. This means that the requests are not completely chronologically ordered in the WAPI usage file: a log entry printed after another one, may have been submitted earlier, thus it should have an earlier timestamp. As the difference in this case may be in milliseconds, if the timestamps are not logged precisely enough, the log entries may appear with the same timestamp (as in the example showed in Table \ref{timestamps}), making their ordering unreliable \cite{bose2013}. This was the case of the logs from MSF. The timestamp granularity was in seconds, thus several requests had the same timestamp. We tried to fix this issue by keeping at least the order that the requests had in the file, and for every two consecutive request with the same timestamp, we added one millisecond to the latter one. 
    
    \begin{table*}
    \vspace{-10pt}
     \caption{Log entries with wrong order because of coarse timestamp. }
        \begin{center}
      \begin{tabular}{ |c|c|c|c| } 
        \hline
         \makecell{Client IP} & \makecell{Request} & \makecell{Timestamp \\(as appears in the logs)} & \makecell{The real time} \\ 
         \hline
         IP1 & request1 & 18/Dec/2020 & 18/Dec/2020 16:05:55.824 \\ 
         \hline
        IP1 & request3 & 18/Dec/2020 & 18/Dec/2020 16:05:55.912 \\  
         \hline
         IP1 & request2 & 18/Dec/2020 & 18/Dec/2020 16:05:55.859 \\ 
         \hline
        \end{tabular}
        \end{center}
          \vspace{-20pt}
        \label{timestamps}
        \end{table*}
        
\textit{Impact}: Having the requests in wrong order may adversely affect analyses' results, by producing erroneous usage patterns, and also hiding important ones.  
     
\textit{Mitigation}: To be able to order log entries exactly in a chronological way, providers should log the timestamp with high precision (e.g., milliseconds). 
\end{itemize}
\end{enumerate}

We have summarized in Table \ref{guidelines}, the mitigation suggestions, based on the main problem they aim to solve. Actually, the potential errors derived from them can affect not only the log pre-processing, but also the analyses, resulting in erroneous usage patterns. Thus, to assist providers in enhancing the usage logs of their WAPIs, we introduce this set of suggestions, that will not just help them to remedy the issues' effects on the data, but uncover and mitigate their root causes. 

\begin{table}[htbp]
\vspace{-10pt}
  \caption{Issues' mitigation for a better WAPI usage logging}
  \begin{center}
    \begin{tabular}{|l|l|}
      \hline
      \makecell{{WAPI usage log issue}}& \makecell{{Mitigation}}\\
      \hline
      \multirow{1}{*}{{Field extraction}}&
     \makecell{{Use a machine parse-able format for logs}}\\
      \hline
      \multirow{4}{*}{{Session Identification}}&
      \makecell{{Provide application identifiers}}\\ \cline{2-2}  &
      \makecell{{Provide different application identifiers for development phase}}\\  \cline{2-2}  &
      \makecell{{Log the referer, user agent}}\\ 
       \cline{2-2}  &
      \makecell{{Log the timestamp in high precision}}\\
      \hline
    \end{tabular}
  \end{center}
  \label{guidelines}
   \vspace{-30pt}
\end{table}

\section{Conclusion and future work} \label{conclusion}

In this paper, we first show the potential of WAPI usage logs, by describing several analyses that providers may perform. Since the success of analyses strongly depends on the quality of their input data, we report the main issues of WAPI usage logs. We then conduct a case study to show the importance of the right log format. Next, derived from the case study, we identify a set of issues that may be present to these logs, explain how these issues impact the analyses, and suggest how to mitigate them.

Our results indicate that WAPI usage logs contain invaluable information about consumers' behavior, needs and difficulties. But this beneficial information comes at the cost of the logs tedious pre-processing. Typically, WAPI usage logs suffer from several issues, that should be properly addressed or mitigated, in order for them to be further analyzed. While some of these issues are related to the nature of the communication between WAPI and its consumers, others may occur because of improper logging. Furthermore, there are many demanding analyses, whose requirements should drive providers in the way they log the usage of their WAPIs.

As a future work, we plan to perform the proposed analyses on the WAPI usage logs, applying first the suggestions in mitigating the existing quality issues.

\end{document}